\documentclass[12pt,preprint]{aastex}

\def\etal{{\it et~al.~}}
\def\nt{{nonthermal~}}
\def\bsax{{\it BeppoSAX~}}

\def\ginga{{\it Ginga~}}

\def\rxte{{\it RXTE~}}

\def\cts{~{\rm counts~s}$^{-1}$~}

\begin{document}

\newcommand{\lessim}{\ \raise -2.truept\hbox{\rlap{\hbox{$\sim$}}\raise5.truept
    \hbox{$<$}\ }}

\title{Nonthermal hard X-ray excess in the Coma cluster: resolving the discrepancy
between the results of different PDS data analyses}

\author{Roberto Fusco-Femiano$^a$, Raffaella Landi$^b$, Mauro Orlandini$^b$
\footnote{$^a$Istituto di Astrofisica Spaziale e Fisica Cosmica
(IASF/Roma), INAF, via del Fosso del Cavaliere, I--00133 Roma,
Italy - roberto.fuscofemiano@iasf-roma.inaf.it; $^b$IASF/Bologna,
INAF, via Gobetti 101, I--40129 Bologna, Italy -
landi@iasfbo.inaf.it - orlandini@iasfbo.inaf.it}}


\affil{}

\begin{abstract}
The detection of a \nt excess in the Coma cluster spectrum by two
\bsax observations analyzed with the XAS package (Fusco-Femiano
\etal) has been disavowed by an analysis (Rossetti \& Molendi)
performed with a different software package (SAXDAS) for the
extraction of the spectrum. To resolve this discrepancy we
reanalyze the PDS data considering the same software used by
Rossetti \& Molendi. A correct selection of the data and the
exclusion of contaminating sources in the background determination
show that also the SAXDAS analysis reports a \nt excess with
respect to the thermal emission at about the same confidence level
of that obtained with the XAS package ($\sim 4.8\sigma$). Besides,
we report the lack of the systematic errors investigated by
Rossetti \& Molendi and Nevalainen \etal taking into account the
whole sample of the PDS observations off the Galactic plane, as
already shown in our data analysis of Abell 2256 (Fusco-Femiano,
Landi \& Orlandini). All this eliminates any ambiguity and
confirms the presence of a hard tail in the spectrum of the Coma
cluster.

\end{abstract}


\section{Introduction}

Nonthermal hard X-ray (HXR) emission was predicted in the
seventies in clusters of galaxies showing extended radio regions,
radio halos or relics, since the same radio synchrotron electrons
can interact with the CMB photons to give inverse Compton (IC)
X-ray radiation (Perola \& Reinhardt 1972; Rephaeli 1979). Several
attempts to detect hard tails in the spectrum of a few clusters of
galaxies were performed with various experiments (Bazzano \etal
1984,90; Rephaeli, Gruber \& Rothschild 1987; Rephaeli \& Gruber
1988; Rephaeli, Ulmer \& Gruber 1994) that reported only upper
limits to the \nt flux. A significant breakthrough in the
measurement of \nt HXR emission was obtained thanks to the
improved sensitivity and wide spectral capabilities of the \bsax
and Rossi X-Ray Timing Explorer (\rxte) satellities. As pointed
out by Petrosian (2003), the discovery of \nt HXR radiation has
led to a remarkable increase of the theoretical investigations
regarding the possible acceleration mechanisms and origin of the
relativistic electrons responsible for the \nt emission, although
the presence of \nt phenomena in the intracluster medium (ICM) of
some clusters was established decades ago (Willson 1970).

Nonthermal HXR radiation was detected in excess of the thermal
emission in the Coma cluster by a first \bsax observation
(Fusco-Femiano \etal 1999) using the Phoswich Detection System
(PDS) and confirmed by a second independent observation with a
time interval of about 3 yr (Fusco-Femiano \etal 2004; thereafter
FF04). The presence of a second component in the X-ray spectrum of
the cluster has been reported also by two \rxte observations
(Rephaeli, Gruber \& Blanco 1999; Rephaeli \& Gruber 2002). A2256
is the second cluster where a \nt excess has been measured by two
\bsax observations (Fusco-Femiano \etal 2000; Fusco-Femiano \etal
2005) and by \rxte (Rephaeli \& Gruber 2003). At a lower
confidence level, with respect to Coma and A2256, \nt HXR
radiation has been detected by \bsax in A754 (Fusco-Femiano \etal
2003). An upper limit to the \nt flux has been reported in A3667
(Fusco-Femiano \etal 2001), A119 (Fusco-Femiano \etal 2003) and
A2163 (Feretti \etal 2001). For the last cluster a \rxte
observation shows instead the presence of a HXR excess (Rephaeli,
Gruber, \& Arieli 2006). \rxte reports also some evidence of \nt
emission by the Bullet Cluster (Petrosian, Madejski, \& Luli
2006).

The PDS spectra of all the \bsax observations were extracted using
the XAS version 2.1 package (Chiappetti \& Dal Fiume 1997)
specifically created to handle the PDS peculiarities. However, a
PDS data analysis performed with a different software package
(SAXDAS) does not report evidence for the presence of a hard tail
in spectrum of the Coma cluster (Rossetti \& Molendi 2004;
thereafter RM04).

To solve this contradiction we reanalyzed the PDS data of the two
\bsax observations using the SAXDAS 2.0.2 package and in this
letter we present our results and conclusions. In Sect. 2 we show
the results of this analysis that are compared with those reported
in FF04 using the XAS package and we examine the possible
systematic errors investigated by RM04 and Nevalainen \etal (2004,
thereafter NE04). In Sec. 3 we discuss the reasons that have led
to controversial results using different software packages.
Finally, Sect. 4 is devoted to the conclusions regarding the
presence of a hard tail in the spectrum of the Coma cluster.

Throughout this Letter we assume a $\Lambda$CDM cosmology with
$H_o$ = 70~km~s$^{-1}$~Mpc$^{-1}$~$h_{70}$, $\Omega_m$ = 0.3 and
$\Omega_{\Lambda}$ = 0.7. An angular distance of $1^{\prime}$
corresponds to 29.0 kpc ($z_{Coma}$ = 0.0232). Quoted confidence
intervals are at $90\%$ level, if not otherwise specified.

\section{PDS Data Reduction and Results}

\begin{figure}
\centering \rotatebox{-90}{ \epsscale{0.8} \plotone{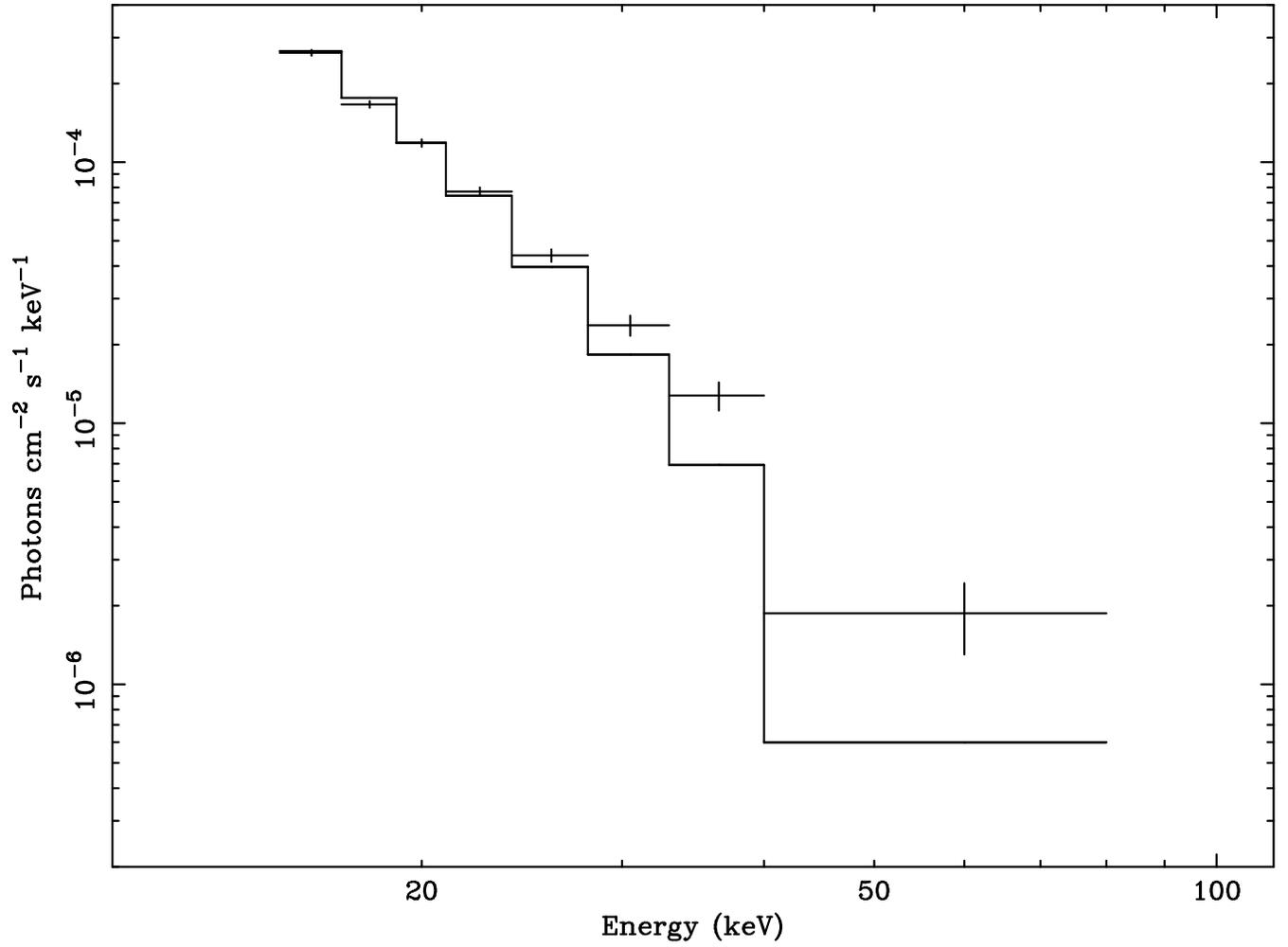}}
\caption{PDS combined spectrum of the Coma cluster obtained with
the XAS package (FF04). The continuous line represents the best
fit with a thermal component at the average cluster gas
temperature of 8.11 keV (David \etal 1993). The error bars are
quoted at the 1$\sigma$ level. The spectrum starts at 15 keV.
\label{fig1}}
\end{figure}

The Coma cluster was observed for the first time by the PDS
instrument (Frontera \etal 1997) in December 1997 for $\sim$91 ks
and re-observed in December 2000 for $\sim$300 ks. The pointing
coordinates of \bsax are at J(2000): $\alpha:~12^h~ 58^m~ 52^s$;
$\delta:~ +27^{\circ}~ 58'~ 54''$. To resolve the discrepancy
between the results shown by FF04 and RM04 regarding the presence
of a \nt component in the HXR Coma spectrum, we have re-analyzed
the PDS data with the SAXDAS package used by RM04. However, we
start by reporting the procedure used in the XAS analysis and the
main results obtained. More details can be found in FF04.

\subsection{XAS analysis}

The PDS data used in the XAS analysis were selected by a first
automatic procedure followed by a visual check in order to
eliminate all the remaining spikes that can affect significantly
the analysis results. The effective exposure times in the two PDS
observations were 44.5 ks and 122.2 ks, respectively (thereafter
OBS1 and OBS2). Since the source is rather faint in the PDS band
($\sim$5 mCrab in 15--100 keV) a careful check of the background
subtraction was performed, making use of the default rocking law
of the two PDS collimators that samples ON/+OFF, ON/--OFF fields
for each collimator with a dwell time of 96 sec (Frontera \etal
1997). When one collimator is pointing ON source, the other
collimator is pointing toward one of the two OFF positions.
Initially, we used the standard procedure to obtain PDS spectra
(Dal~Fiume \etal 1997); this procedure consists of extracting one
accumulated spectrum for each unit for each collimator position.
We then checked the two independently accumulated background
spectra in the two different +/--OFF sky directions, offset by
$210'$ with respect to the on-axis pointing direction (+OFF
pointing: $\alpha:~12^h~58^m~ 57.8^s$; $\delta:~ +24^\circ~ 28'~
55''.1$ --OFF pointing: $\alpha:~12^h~58^m~ 47.0^s$; $\delta:~
+31^\circ~ 28'~ 54''.7$). The comparison between the two
accumulated backgrounds (difference between the +OFF and --OFF
count rate spectra) showed that for OBS1 the difference was
compatible with zero ($0.044\pm 0.047$ \cts for a background level
of $21.66\pm 0.02$ \cts in 15--100 keV), while for the longer,
more sensitive OBS2, there was an excess of $0.064\pm 0.021$ \cts
(background $16.76\pm 0.01$ \cts)\footnote{The $\sim$20\%
variation in the PDS background is due to the \bsax orbital decay:
the lower orbit for OBS2 increased the shielding to ambient
particles, therefore lowering the diffuse background.}. As
reported in FF04, a careful check of possible variable sources in
the PDS offset fields led the attention to the BL~Lac source
1ES~1255+244, present in the +OFF field, that was observed by
\bsax on May 1998 in the framework of a spectral survey of BL~Lacs
by Beckmann \etal (2002). Because of the very short exposure time
($\sim$ 3~ksec), our analysis of the source has determined only a
2$\sigma$ upper limit of 0.26 \cts in 15--100 keV, corresponding
to 1.6 mCrab, however compatible with the background excess
measured in OBS2. Moreover, just in the center of the +OFF field
is also present the extremely weak ROSAT source
RX~J125847.1+242741. The presence of these contaminating sources
justified the decision to exclude the +OFF field in the background
evaluation and consider only the --OFF field as the
"uncontaminated" background for both the Coma observations.

The combined spectrum, obtained by summing the spectra of the two
observations (OBS1 \& OBS2), shows a \nt hard excess with respect
to the thermal component at the confidence level (c.l.) of $\sim
4.8\sigma$ in the 20-80 keV energy range considering only the -OFF
background spectrum. The spectrum is reported in Fig.1. FF04 also
reported the confidence level of the \nt excess considering the
standard procedure, i.e. by using as background the average of the
spectra extracted from both the two offset fields. The c.l. is
lower ($\sim 3.9\sigma$) due to the contamination in the +OFF
background. The average gas temperature used in the analysis is
that measured by \ginga of 8.11$\pm$0.07 keV (David \etal 1993) in
a field of view comparable to that of the PDS (FWHM $\sim
1.3^{\circ}$). This value of the average temperature is confirmed
by a determination of \rxte that reports a best-fit temperature of
7.90$\pm$0.03 keV (Rephaeli \& Gruber 2002) in a field of view of
$\sim 1^{\circ}$ comparable to the field of view of \ginga and
PDS.

\subsection{SAXDAS analysis}

The first combined spectrum was extracted considering the PDS data
resulting by the automatic selection operated by the SAXDAS
package \footnote{See
http://www.asdc.asi.it/bepposax/software/saxdas/}. The confidence
level of the excess, in the range 20-80 keV and for an average gas
temperature of 8.11 keV, is $\sim$ 2.9$\sigma$ (observed count
rate = 0.1717$\pm$0.0146 \cts, model predicted rate = 0.1295 \cts)
taking into account only the uncontaminated -OFF position. The
effective exposure time is 42.8 ks in OBS1 and 119.3 ks in OBS2
for a total exposure time of 162.1 ks. To resolve the discrepancy
between these results and those obtained by FF04 with a XAS
analysis (see previous section), a new combined spectrum was
extracted considering the \textit{same} time windows used in the
XAS analysis. These time windows are different from those used by
RM04 in their SAXDAS analysis. The reason is in the different
criterion for eliminating the spurious spikes produced by the
charged particles that interact with the PDS detectors: a
semi-automated procedure (i.e. an automated procedure followed by
a visual check) for XAS and a completely automated procedure for
SAXDAS (Guainazzi \& Matteuzzi 1997). With the same time windows
used by XAS, the c.l. of the excess in the SAXDAS spectrum raises
at $\sim$ 4.2$\sigma$ (observed count rate = 0.1902$\pm$0.0148
\cts, model predicted rate = 0.1280 \cts), while the total
exposure time reduces to 160.9 ks (it was 162.1 ks).

\begin{figure}
\centering \rotatebox{-90}{ \epsscale{0.8} \plotone{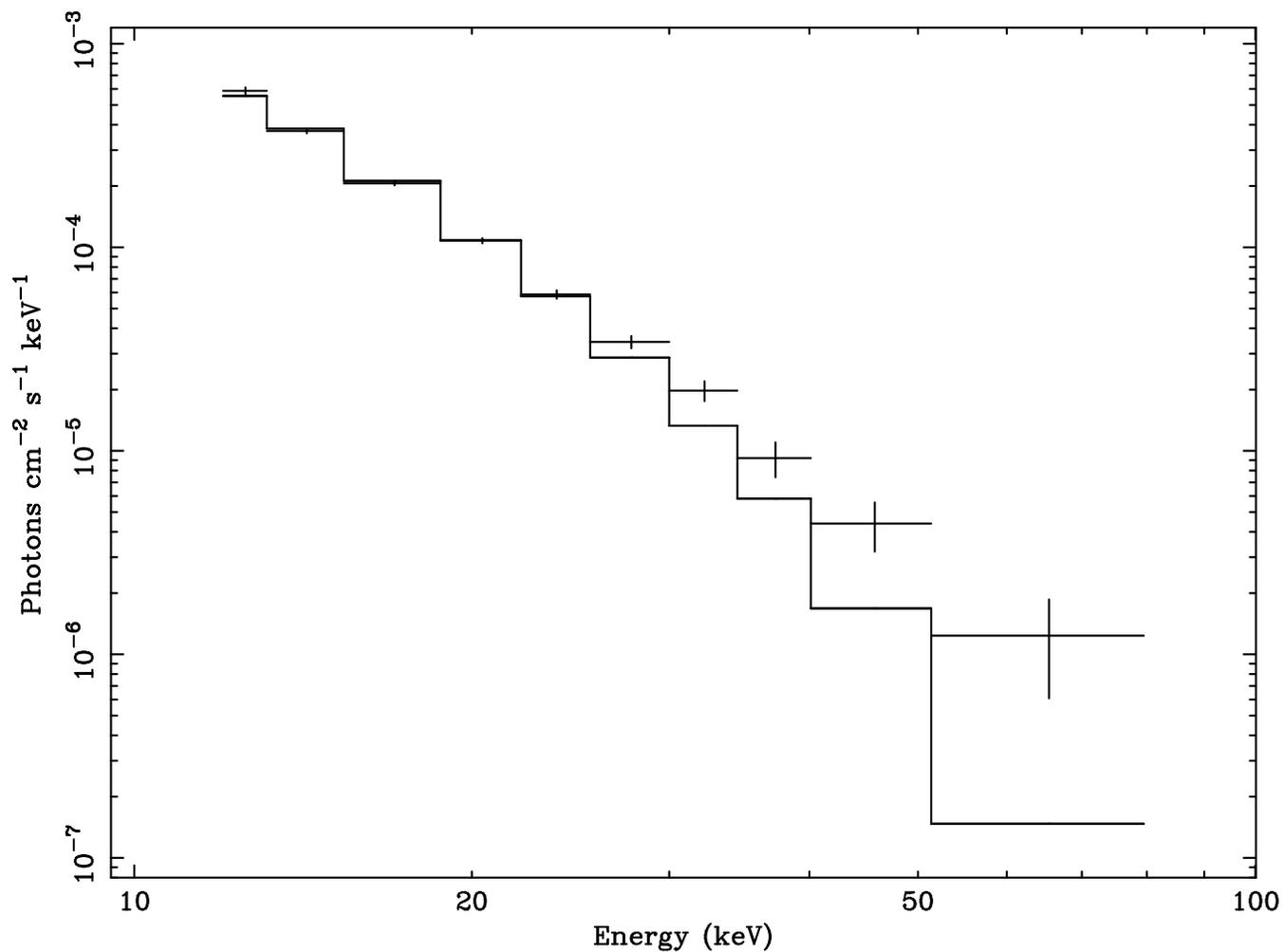}}
\caption{PDS combined spectrum of the Coma cluster obtained with
the SAXDAS package (see text). The continuous line represents the
best fit with a thermal component at the same average cluster gas
temperature of Fig. 1. The error bars are quoted at the 1$\sigma$
level. The spectrum starts at 12 keV.\label{fig2}}
\end{figure}

The two packages, starting from the same time windows, produce a
difference in the total effective exposure time of 5.8 ks
($t_{XAS}$ = 166.7 ks, $t_{SAXDAS}$ = 160.9) essentially due to:
a) the different earth angle above which data are selected (the
earth angle is defined as the angle between the earth limb and the
\bsax pointing direction). The earth angle is 5$^{\circ}$ for XAS
and 10$^{\circ}$ for SAXDAS; b) the removal of observational time
after any South Atlantic Geomagnetic Anomaly (SAGA) passage. The
SAXDAS package removes 5 minutes after any passage, while XAS
eliminates the time necessary to reach the correct voltage of the
instruments after their shut-down during the SAGA passage.

In conclusion, we have a lower exposure time for the SAXDAS
analysis which may imply a lower c.l. of the \nt excess. To
quantify this difference we have extracted a further combined
spectrum using the same time windows selected in the XAS analysis
but imposing an earth angle of 5$^{\circ}$ in the SAXDAS package.
The total effective time exposure of the two observations is now
$t_{SAXDAS}$ = 169.1 ks instead of 160.9 ks. The excess results to
be at the c.l. of $\sim$ 4.6$\sigma$ (observed count rate =
0.1944$\pm$0.0144 \cts, model predicted rate = 0.1280 \cts). This
HXR spectrum is reported in Fig. 2. The difference in the initial
energy of the SAXDAS and XAS spectra (SAXDAS spectrum starts at 12
keV, while XAS at 15 keV) gives insignificant variations in the
c.l. values of the \nt excess.

\subsection{Possible systematic errors}

The systematic errors examined by RM04 and NE04 have already been
discussed by us in the PDS analysis of A2256 (Fusco-Femiano, Landi
\& Orlandini 2005), but considering their importance in the
analysis results we intend to report here again the main parts of
the discussion.

1) RM04 claim the presence of an "Instrumental background
residual" (see sect. 2.1 of their paper) derived from the analysis
of 15 "blank fields", i.e. fields which do not contain sources
showing significant emission in the PDS energy range. By summing
the spectra from these observations they find that the spectrum
differs from zero: the count rate is ($1.45\pm 0.77)\times
10^{-2}$ \cts in the 12-100 keV energy range. This seems to
indicate that the background in the ON position is larger than
that in the $\pm$OFF positions producing an instrumental
contribution not removed by the background subtraction procedure.
This effect has been studied by Landi \etal (2005) considering the
complete sample of 868 PDS pointings with galactic latitude
$|b|>15^\circ$, and selecting the 15--100 keV net count spectra
for which there is source detection below 1$\sigma$ (that is,
``blank fields''). These spectra have been summed imposing a net
exposure greater than 20~ksec. A net count rate of $(1.67\pm
5.30)\times 10^{-3}$ \cts has been derived, consistent with the
definition of ``blank field''. Also NE04 do not report evidence
for an instrumental residual.

2) The other effect evaluated by RM04 regards the systematic
differences between the background fields. They analyze a sample
of 69 observations whose target is outside the galactic plane and
with a long exposure time (see Appendix of their paper). RM04 find
that the mean value of the difference between ON and the two -OFF
and +OFF sky positions is significantly different from zero and
positive. Also this effect has been investigated by Landi \etal
(2005) on the {\em whole} sample of PDS observations. The obtained
value of $(5.3\pm 6.3)\times 10^{-3}$ \cts is consistent with no
contamination at all. We presume that the value found by RM04
could be due to the small sample of observations they considered.
NE04, analyzing a larger sample of data with respect to that used
by RM04 (164 PDS observations), found a systematic difference
between ON and the two offset pointings that cancels out in the
standard usage of both offsets.

3) In addition, NE04 introduce a systematic error in the net
source count rates due to unresolved and not significantly
detected point sources present in the PDS field of view. They find
that an excess of 0.019 counts~s$^{-1}$ has to be added to the net
count rate spectra (no errors are given on this measurement), when
the standard method of background evaluation is used, and 0.027
counts~s$^{-1}$ has to be added when the background is evaluated
from only one offset field. We performed this same analysis on a
set of 868 observations (NE04 use 164 fields) and find that the
contribution of background fluctuations due to unresolved and not
significantly detected sources in the offset fields (in other
words, the PDS confusion limit) introduces a variance
$\sigma^2_{fluc} = (9.5\pm 10.3)\times 10^{-4}$
(counts~s$^{-1}$)$^2$ consistent with zero (Fusco-Femiano, Landi
\& Orlandini 2005). Therefore PDS data are not affected by this
systematic effect.

\section{Discussion}

Fig. 2 shows that also the PDS combined spectrum extracted with
the SAXDAS package evidences the presence of a \nt excess with
respect to the thermal emission at about the same c.l. reported by
the XAS analysis (FF04) ($\sim$ 4.8$\sigma$ for XAS and $\sim$
4.6$\sigma$ for SAXDAS). The discrepancy between the results
reported in FF04 and RM04 is mainly due: a) to an accurate
selection of the events. In fact, we have shown that the SAXDAS
analysis of the same PDS time windows used with the XAS software
leads to a significant increase of the c.l. of the excess (from
$\sim 2.9\sigma$ to $\sim 4.2\sigma$); b) to a correct
determination of the background. The XAS and SAXDAS spectra shown
in Fig.1 and Fig. 2, respectively, are obtained considering only
the -OFF background. The exclusion of the +OFF position in the
background determination implies a more pronounced detection of
\nt radiation from the Coma cluster. The check regarding the
presence of contaminating sources in the +OFF field was not
operated by RM04.

An increase of $\sim$ 0.4$\sigma$ in the c.l. value is obtained
adopting in the SAXDAS analysis the same earth angle of
$5^{\circ}$ used in the XAS procedure. The lower earth angle
implies a greater exposure time that justify an improvement in the
c.l. value of the \nt excess. Considering in the SAXDAS analysis
the same time windows and the same earth angle used in the XAS
analysis the difference in the total exposure time is 2.4 ks in
favour of the SAXDAS package ($t_{XAS}$ = 166.7 ks, $t_{SAXDAS}$ =
169.1 ks) due to the different removal of the time operated by the
two packages after any SAGA passage. The XAS package results to be
more conservative waiting more time than SAXDAS to allow the PDS
high voltage to reach the correct levels. So, some spurious events
could be present in the SAXDAS analysis, in particular for the
longer OBS2 observation.

In order to reproduce the results obtained by RM04 we have
computed the c.l. of the excess for OBS1 and OBS2 spectra (RM04 do
not report the c.l. value for the combined spectrum) using the
automatic selection of the events operated by the SAXDAS package
and the standard procedure for the background determination. The
c.l. values are 2.90$\sigma$ and 1.34$\sigma$ for OBS1 and OBS2,
respectively, in the energy range 25-80 keV and for an average gas
temperature of 8.21 keV. These values are not much distant from
those reported in Table 2 of RM04 (2.84$\sigma$ for OBS1 and
1.11$\sigma$ for OBS2)\footnote{These values of the c.l. of the
excess refer to the line "No Subtraction" (no subtraction of the
instrumental residual that Landi \etal (2005) find consistent with
zero; see Sect. 2.3).} confirming that the selection procedure of
the PDS events and the exclusion of the +OFF position for the
background determination are the main reasons of the different
results reported in FF04 and RM04. The PDS data analysis of very
weak sources like Coma in the HXR band requires a rigorous
selection of the events in order to eliminate the presence of any
spikes able to introduce \textit{noise} that hides the presence of
a \nt excess with respect to the thermal radiation. The presence
of contaminating sources in the offset fields does not allow the
use of the standard procedure for the background evaluation.

\section{Conclusions}

We have shown that the presence of a \nt excess with respect to
the thermal emission in the spectrum of the Coma cluster does not
depend on the used software package (XAS or SAXDAS) for the PDS
data analysis. The spectra extracted with XAS and SAXDAS, reported
in Fig. 1 and 2, respectively, show both a \nt excess at about the
same c.l. value when in the SAXDAS analysis the same time windows
used in the XAS analysis (FF04) are adopted and the +OFF sky
direction is not taken into account in the background
determination for the presence of contaminating sources. The
systematic effects claimed by RM04 and NE04 can be excluded
considering the whole sample of the PDS observations off the
Galactic plane.

This re-analysis of the PDS data, using the SAXDAS package,
explains the different results reported in FF04 and RM04
confirming the presence of a \nt component in the Coma cluster
spectrum.

\section{Acknowledgments}
We wish to thank L.Feretti and G.Giovannini for a critical reading
of the paper and the referee for his helpful comments on the
manuscript.

\clearpage


\newpage
\begin{references}

\reference{}Bazzano, A., Fusco-Femiano, R., La Padula, C.,
Polcaro, V.F., \& Ubertini, P. 1984, ApJ, 279, 515

\reference{}Bazzano, A. \etal 1990, ApJ, 362, L51

\reference{}Beckmann, V. \etal 2002, A\&A 383, 410

\reference{}Chiappetti, L., \& Dal Fiume, 1997, in Data Analysis
in Astronomy, ed. V.Di Gesu` \etal (Singapore: World Scientific),
101

\reference{}Dal Fiume, D. \etal 1997, Proc. of the Fifth Workshop
{\it "Data Analysis in Astronomy"}, eds.: V. Di Gesu' \etal , p.
111

\reference{}David, L.P., Slyz, A., Jones, C., Forman, W., \&
Vrtilek, S.D. 1993, ApJ, 412, 479

\reference{}Feretti, L., Fusco-Femiano, R., Giovannini, G., \&
Govoni, F. 2001, A\&A, 373, 106

\reference{}Frontera, F., Costa, E., Dal~Fiume, D., Feroci, M.,
Nicastro, L., Orlandini, M., Palazzi, E., \& Zavattini, G. 1997,
A\&AS., 122, 357

\reference{}Fusco-Femiano, R., Dal Fiume, D., Feretti, L.,
Giovannini, G., Grandi, P., Matt, G., Molendi, S., \& Santangelo,
A. 1999, ApJ, 513, L21

\reference{}Fusco-Femiano, R., Dal Fiume, D., De Grandi, S.,
Feretti, L., Giovannini, G., Grandi, P., Malizia, A., Matt, G., \&
Molendi, S. 2000, ApJ, 534, L7

\reference{}Fusco-Femiano, R., Dal Fiume, D., Orlandini, M.,
Brunetti, G., Feretti, L., \& Giovannini, G. 2001, ApJ., 552, L97

\reference{}Fusco-Femiano, R. \etal 2003, A\&A, 398, 441

\reference{}Fusco-Femiano, R. \etal 2003, Proc. of "Matter and
Energy in Clusters of Galaxies", Taiwan, ASP Conf.Ser., eds.:
S.Bowyer and C.-Y. Hwang; p. 109

\reference{}Fusco-Femiano, R., Orlandini, M., Brunetti, G.,
Feretti, L., Giovannini, G., Grandi, P., \& Setti, G. 2004, ApJ,
602, L73 (FF04)

\reference{}Fusco-Femiano, R., Landi, R., \& Orlandini, M. 2005,
ApJ, 624, L69

\reference{}Guainazzi, M. \& Matteuzzi, A. 1997, SDC Technical
Report - TR-
011\\ftp://www.asdc.asi.it/pub/sax/doc/reports/sdc-tr14.ps.gz

\reference{}Landi, R. 2005, Ph.D. thesis, Bologna Univ. (paper in
preparation by Orlandini \etal)

\reference{}Nevalainen, J., Oosterbroek, T., Bonamente, M., \&
Colafrancesco, S. 2004, ApJ, 608, 166 (NE04)

\reference{}Perola, G.C., \& Reinhardt, M. 1972, A\&A, 17, 432

\reference{}Petrosian, V. 2003, ASPC, 301, 337

\reference{}Petrosian, V., Madejski, G., \& Luli, K. 2006,
accepted by ApJ (astro-ph/0608455)

\reference{}Rephaeli, Y. 1979, ApJ, 227, 364

\reference{}Rephaeli, Y., Gruber, D.E., \& Rothschild, R.E. 1987,
ApJ, 320, 139

\reference{}Rephaeli, Y. \& Gruber, D.E. 1988, ApJ, 333, 133

\reference{}Rephaeli, Y., Ulmer, M., \& Gruber, D.E. 1994, ApJ,
429, 554

\reference{}Rephaeli, Y., Gruber, D.E., \& Blanco, P. 1999, ApJ,
511, L21

\reference{}Rephaeli, Y., \& Gruber 2002, ApJ, 579, 587

\reference{}Rephaeli, Y., \& Gruber 2003, ApJ, 595, 137

\reference{}Rephaeli, Y., Gruber, D., \& Arieli, Y. 2006, accepted
by ApJ (astro-ph/0606097)

\reference{}Rossetti, M., \& Molendi, S. 2004, A\&A, 414, L41
(RM04)

\reference{}Willson, M.A.G. 1970, MNRAS, 151, 1

\end{references}
\end{document}